\documentclass[11pt]{article}
\usepackage{hyperref}
\pdfoutput=1
\begin{document}
\title{Liquid bridge pinch off and satelite drop formation under thermocapillary effect in Japanese Experiment Module ‘Kibo’ aboard the International Space Station}
\author{Ichiro UENO \and Koichi NISHINO \and Mitsuru OHNISHI \and Hiroshi KAWAMURA \and Masato SAKURAI \and Satoshi MATSUMOTO\\
\\\vspace{6pt} \\Tokyo University of Science, Noda, Chiba 278-8510, Japan
\\Yokohama National University, Yokohama, Kanagawa 240-8501, Japan
\\Tokyo University of Science, Suwa, Chino, Nagano 391-8502, Japan
\\Japan Aerospace Exploration Agency, Tsukuba, Ibaraki 305-8505, Japan
\\Japan Aerospace Exploration Agency, Chofu, Tokyo 182-8522, Japan}

\maketitle

\begin{abstract}
The long-duration fluid physics experiments on a thermocapillary-driven flow have been carried out on the Japanese experiment module ‘Kibo’ aboard the International Space Station (ISS) since 2008. In these experiments, various aspects of thermocapillary convection in a half-zone (HZ) liquid bridge of high Prandtl number fluid have been examined under the advantages of the long-duration high-quality microgravity environment. This fluid dynamics video introduce a pinch off of liquid bridge of 30 mm in diameter as a part of the on-orbit experiments. The effect of thermocapillary-driven flow on the pinch off and satellite drop formation is examined.
\end{abstract}

\section{}
Operation of the experimental facilities in the Japanese Experimental Module, known as ‘Kibo,’ that means ‘hope’ in Japanese, aboard the International Space Station (ISS) has started since 2008. Marangoni Experiment In Space (MEIS) as a fluid physics experiment has conducted as the first scientific experiment in the Kibo. This project consists of five series of the experiments from MEIS-I to -V; the MEIS-I was carried out in 2008, and -II in 2009, -IV in 2010 and -III in 2011. The target phenomenon of this project is the thermocapillary-driven flow, which comes from a temperature dependence of the surface tension.\\
Such a phenomenon becomes obvious under a microgravity condition and/or a micro-scale condition, in which the body force becomes too small comparing to the surface force, so that this has attracted scientists and engineers for crystal growth processes, chemical reaction controls, handling techniques of fluids with free surfaces, and so on, in microgravity environments.\\
The target geometry is the half-zone (HZ) liquid bridge, which has been employed for fundamental studies on the floating zone (FZ) method for the single crystal growth of semiconductors. In the HZ method, a liquid is suspended between two parallel disks to form the liquid bridge. One disk is heated and the other is cooled to realize a designated temperature difference between the disks. Such a temperature difference causes a non-uniform temperature distribution over the free surface to result in a thermocapillary-drive flow. In the present study, we have focused on the non-linear flow regimes induced in the liquid bridge and in the hanging droplet. We have also focused on the pinch off of the relatively large-scale liquid bridge.\\
Experimental conditions are as follows; silicone oil of 10 cSt in kinematic viscosity is employed as the test fluid. A liquid bridge if formed between the coaxial cylindrical rods of 30 mm in diameter. One rod is heated and the other is cooled to realize a designated temperature difference between the end surfaces of the liquid bridge. A fine bellows is installed to the rod for cooling to realize a control of the volume of the liquid bridge at a designate flow rate. In the pinch off experiments, we form the liquid bridge with a designated aspect ratio and a volume ratio. Then small enough volume of the fluid is sucked by the fine bellows, and the liquid bridge is left for a while to wait for a starting of spontaneous pinch off.
This experimental facility consists of several measurement equipments; infrared camera for measuring the surface temperature of the liquid bridge, thermocouple for measuring a local temperature near the free surface of the liquid bridge, CCD cameras for observing the liquid bridge and suspended particles from the top and the side.\\
This movie consists of four major parts; (1) liquid bridge formation process by telecommunication procedures, liquid bridge pinch off (2) under isothermal and (3) non-isothermal conditions, and (4) comparison of the phenomena themselves and numerical predictions conducted by one of the authors (MO).\\
\end{document}